\documentclass{PoS}
\usepackage{cite}

\title{Quark orbital angular momentum in the proton evaluated using a
direct derivative method}

\ShortTitle{Quark orbital angular momentum in the proton}

\author{\speaker{M.~Engelhardt}$\,\,^{a\, \dagger} $, J.~Green$^{b}$,
N.~Hasan$^{c,d}$, S.~Krieg$^{c,d}$, S.~Meinel$^{e,f}$, J.~Negele$^{g}$,
A.~Pochinsky$^{g}$ and S.~Syritsyn$^{f,h}$ \\
        $^{a} $Department of Physics, New Mexico State University, Las Cruces,
        NM 88003, USA\\
        $^{b} $NIC, Deutsches Elektronen-Synchroton, 15738 Zeuthen, Germany\\
        $^{c} $Bergische Universit\"at Wuppertal, 42119 Wuppertal, Germany\\
        $^{d} $IAS, J\"ulich Supercomputing Centre, Forschungszentrum
        J\"ulich, 52425 J\"ulich, Germany\\
        $^{e} $Department of Physics, University of Arizona, Tucson, AZ 85721,
        USA\\
        $^{f} $RIKEN BNL Research Center, Brookhaven National Laboratory,
        Upton, NY 11973, USA\\
        $^{g} $Center for Theoretical Physics, Massachusetts Institute of
        Technology, Cambridge, MA 02139,\\ $\mbox{\hspace{0.02cm} } $ USA\\
        $^{h} $Department of Physics and Astronomy, Stony Brook University,
        Stony Brook, NY 11794, USA\\
        $^{\dagger} $E-mail: \email{engel@nmsu.edu}}

\abstract{Quark orbital angular momentum (OAM) in the proton can be
calculated directly given a Wigner function encoding the simultaneous
distribution of quark transverse positions and momenta. This
distribution can be accessed via proton matrix elements of a quark
bilocal operator (the separation in which is Fourier conjugate to
the quark momentum) featuring a momentum transfer (which is Fourier
conjugate to the quark position). To generate the weighting by quark
transverse position needed to calculate OAM, a derivative with respect
to momentum transfer is consequently required. This derivative is
evaluated using a direct derivative method, i.e., a method in which
the momentum derivative of a correlator is directly sampled in the
lattice calculation, as opposed to extracting it a posteriori from
the numerical correlator data. The method removes the bias stemming
from estimating the derivative a posteriori that was seen to afflict a
previous exploratory calculation. Data for Ji OAM generated on a clover
ensemble at pion mass $m_{\pi } = 317\, \mbox{MeV} $ are seen to agree
with the result obtained via the traditional Ji sum rule method. By
varying the gauge connection in the quark bilocal operator, also
Jaffe-Manohar OAM is extracted, and seen to be enhanced significantly
compared to Ji OAM.}

\FullConference{23rd International Spin Physics Symposium - SPIN2018 -\\
		10-14 September, 2018\\
		Ferrara, Italy}

\begin{document}

\section{Introduction}
A prominent endeavor in the study of hadron structure is understanding
the decomposition of the spin of the proton into the contributions from
the spins and orbital angular momenta (OAM) of its quark and gluon
constituents. Since gauge invariance prevents a consideration of quark
and gluon degrees of freedom in isolation, there is no unique definition
of quark OAM; any such definition will contain gluonic effects to a varying
degree. Two widely studied decomposition schemes are the ones due to Ji
\cite{jidecomp} and to Jaffe and Manohar \cite{jmdecomp}.

Recently, a method to evaluate quark OAM in Lattice QCD directly from the
simultaneous distribution of quark transverse positions and momenta in a
rapidly propagating proton was explored in \cite{f14pap}. This distribution
is encoded in generalized transverse momentum-dependent parton distributions
(GTMDs) \cite{mms,lorce,eomlir,eomlirl}; in comparison to standard TMDs,
which parametrize forward matrix elements of an appropriate bilocal quark
operator, GTMDs additionally include a momentum transfer. The momentum
transfer is Fourier conjugate to the quark impact parameter and thus
supplements the transverse momentum information with transverse position
information, generating, in effect, a Wigner function. The freedom in
choosing the gauge connection in the bilocal quark operator allows one to
access a continuum of quark OAM definitions, including the ones of Ji and
Jaffe-Manohar. In this respect, the GTMD approach extends beyond the
standard quark OAM calculation via Ji's sum rule, which yields specifically
Ji OAM.

The present work constitutes a further methodological development of the
approach introduced in \cite{f14pap}. The result for Ji OAM obtained
using the concrete implementation in \cite{f14pap} deviated significantly
from the standard Ji sum rule result, due to a systematic bias inherent
in the numerical method. This discrepancy is resolved in the present
work, validating the GTMD approach. Furthermore, the present investigation
is carried out at a significantly lower pion mass,
$m_{\pi } = 317\, \mbox{MeV} $.

\section{Quark orbital angular momentum}
The quark OAM component $L^U_3 $ in a longitudinally
polarized proton propagating in the 3-direction can be accessed via a
GTMD matrix element \cite{lorce},
\begin{equation}
L^U_3 = \frac{1}{2P^+ }
\epsilon_{ij} \frac{\partial }{\partial z_{T,i} }
\left. \frac{\partial }{\partial \Delta_{T,j} }
\frac{\langle p^{\prime } , S^{\prime } =\vec{e}_{3} |
\overline{\psi} (-z/2) \gamma^+ U \psi(z/2)
| p, S=\vec{e}_{3} \rangle }{ {\cal S} [U]}
\right|_{z^+ = z^- =0\, , \ \Delta_{T} =0\, , \ z_T \rightarrow 0}
\label{ldersingle}
\end{equation}
A number of remarks are in order concerning this expression. The initial
and final proton momenta are treated symmetrically,
$p=P-\Delta_{T} /2$, $p^{\prime } =P+\Delta_{T} /2 $, where the
spatial component of $P$ is in 3-direction and the momentum transfer
$\Delta_{T} $ is transverse. Since $\Delta_{T} $ is Fourier conjugate
to the quark impact parameter $b_T $, evaluating the $\Delta_{T} $-derivative
at $\Delta_{T} =0$ amounts to averaging $b_T $. On the other hand, the
transverse quark operator separation $z_T $ is Fourier conjugate to
the transverse quark momentum $k_T $; therefore, evaluating the
$z_T $-derivative at $z_T =0$ amounts to averaging $k_T $. Here, the
limit $z_T \rightarrow 0$ must be taken carefully, since it is
associated with ultraviolet divergences. In aggregate, thus,
(\ref{ldersingle}) yields the average $b_T \times k_T $, i.e.,
OAM in the 3-direction. Also the longitudinal
quark momentum components are integrated over in view of the specification
$z^+ = z^- =0$. In the thus constructed average, quark spin direction
is immaterial owing to the use of the Dirac structure $\gamma^{+} $.
Finally, (\ref{ldersingle}) depends on the gauge link $U$ connecting the
quark operators, along with a soft factor ${\cal S} [U]$ which absorbs
divergences associated with the quantum fluctuations of $U$; for present
purposes, one may consider ${\cal S} [U]$ to include also renormalization
factors associated with the quark field operators. This soft factor is the
same as for the standard TMD matrix element \cite{gtmdsoft}, since
(\ref{ldersingle}) only differs
from the latter in the external state, not the operator. The multiplicative
factor ${\cal S} [U]$ will be canceled by forming an appropriate ratio
below and thus does not need to be specified in more detail. It is in the
path of $U$ that different definitions of quark OAM are encoded;
(\ref{ldersingle}) is a functional of $U$.
In the present work, staple-shaped $U\equiv U[-z/2,\eta v-z/2,\eta v+z/2,z/2]$
are considered, where the arguments of $U$ are positions joined by
straight Wilson lines. Thus, the vector $v$ gives the direction of
the staple, and the length of the staple is scaled by the parameter
$\eta $. For $\eta =0$, one has a straight Wilson line directly connecting
the quark operators.

The $\eta =0$ straight gauge link limit corresponds to Ji OAM \cite{jist},
whereas the $\eta \rightarrow \pm \infty $ limit of a staple extending
to infinity yields Jaffe-Manohar OAM \cite{hatta}. Such a staple link
incorporates final state interactions, e.g., in semi-inclusive
deep inelastic scattering (SIDIS) processes, with the staple legs
corresponding to the direction of propagation of the struck quark.
Thus, Jaffe-Manohar quark OAM differs from Ji quark OAM in that
it includes the integrated torque accumulated by the struck quark as
it leaves the proton \cite{burk}. In a Lattice QCD calculation, $\eta $
can be varied quasi-continuously, with the Jaffe-Manohar limit achieved
by extrapolation. This yields a gauge-invariant interpolation between
the Ji and Jaffe-Manohar cases.

In addition, the direction $v$ of the staple needs to be specified.
The most straightforward choice for the direction of propagation of the
struck quark in a hard scattering process would initially appear to be a 
lightlike vector. However, such a choice leads to severe rapidity
divergences, which are regulated in the scheme advanced in
\cite{aybat,collbook} by taking $v$ off the light cone into the
spacelike region. The matrix element (\ref{ldersingle}) determining quark
OAM therefore depends on the additional Collins-Soper type parameter
$\hat{\zeta } = v\cdot P / (\sqrt{|v^2 |} \sqrt{P^2 } )$.
The light-cone limit corresponds to $\hat{\zeta } \rightarrow \infty $.

As in lattice TMD studies \cite{tmdlat,bmlat,rentmd}, an appropriate ratio
of quantities can be employed to cancel the soft factor ${\cal S} [U]$.
A suitable quantity for this purpose is the number of valence quarks
\begin{equation}
n = \frac{1}{2P^{+} }
\left. \frac{\langle p^{\prime } , S^{\prime } =\vec{e}_{3} |
\overline{\psi}(-z/2) \gamma^+ U \psi(z/2)
| p, S =\vec{e}_{3} \rangle }{ {\cal S} [U]}
\right|_{z^+ =z^- =0\, , \ \Delta_{T} =0\, , \ z_T \rightarrow 0}
\label{ndenom}
\end{equation}
which only differs from (\ref{ldersingle}) by omitting the weighting
with $b_T \times k_T $ (in terms of the Fourier conjugate variables),
and thus counts quarks. The soft factor ${\cal S} [U] $ is even in
$z_T $, and thus cancels when forming the ratio $L^U_3 /n$.
Furthermore, at finite lattice spacing $a$, the derivative with
respect to $z_T $ in (\ref{ldersingle}) is realized as a finite
difference, leading to the renormalized quantity evaluated in
practice,
\begin{equation}
\frac{L^U_3 }{n} = \frac{1}{a} \epsilon_{ij}
\left. \frac{\frac{\partial }{\partial \Delta_{T,j} }
\left( \Phi (a\vec{e}_{i} ) - \Phi (-a\vec{e}_{i} ) \right) }{
\Phi (a\vec{e}_{i} ) + \Phi (-a\vec{e}_{i} )}
\right|_{z^+ = z^- =0\, , \ \Delta_{T} =0}
\label{rdiscrete}
\end{equation}
where summation over the transverse indices $i$ and $j$ is implied, and
the abbreviation
$\Phi (z_T) = \langle p^{\prime } , S^{\prime } =\vec{e}_{3} |
\overline{\psi}(-z/2) \gamma^+ U \psi(z/2) | p, S =\vec{e}_{3} \rangle $
has been introduced. On the other hand, (\ref{rdiscrete}) also calls
for a derivative with respect to $\Delta_{T} $. In the initial exploration
\cite{f14pap}, this derivative was likewise realized as finite difference.
This led to a significant systematic bias in the numerical results because
of the substantial increment in $\Delta_{T} $ employed. The chief advance
of the present study is to evaluate this derivative using a direct
derivative method, as described below. The present study furthermore
is carried out at a lower pion mass, $m_{\pi } = 317\, \mbox{MeV} $,
than used in \cite{f14pap}.

\section{Direct derivative method}
The principal ingredient needed to extract the matrix element $\Phi (z_T) $
is the following three-point correlator, constructed using proton
sources and sinks $\overline{N} $, $N$, projected onto proton momentum
$P+\Delta_{T} /2 $ at the sink and onto momentum transfer $\Delta_{T} $
at the operator insertion, as well as onto longitudinal polarization,
encoded in $\Gamma_{\mbox{\scriptsize pol} } $,
\begin{eqnarray}
C_3 &=&
\mbox{Tr} \left[ \sum_{x,y} e^{-i(P+\Delta_{T} /2)\cdot (x-y)}
e^{-i(P-\Delta_{T} /2)\cdot y}
\left\langle N(x) \overline{\psi}(y-z/2) \gamma^+ U \psi(y+z/2)
\overline{N} (0) \right\rangle \Gamma_{\mbox{\scriptsize pol} }
\right] \label{3ptcorr} \\
&=& \sum_{x,y} e^{-iP\cdot x} \left\langle \mbox{Tr} \left[ \left(
\gamma_{5} G_{\mbox{\scriptsize pt-sm} } (y-z/2,x,\Delta_{T} /2) \gamma_{5}
S^{N\overline{N} \dagger}_{\Gamma_{\mbox{\scriptsize pol} } } (0;x)
\right)^{\dagger } \gamma^{+} U G_{\mbox{\scriptsize pt-sm} }
(y+z/2,0,-\Delta_{T} /2) \right] \right\rangle
\nonumber
\end{eqnarray}
Here, the second line corresponds to the standard evaluation of the
correlator through a sequential source,
$S^{N\overline{N} }_{\Gamma_{\mbox{\scriptsize pol} } } (0;x)$, in which,
however, the phases associated with the projection onto momentum transfer
$\Delta_{T} $ have been absorbed into the propagators,
$G_{\mbox{\scriptsize pt-sm} } (s,t,q) = e^{-iq\cdot(s-t)}
G_{\mbox{\scriptsize pt-sm} } (s,t)$,
where $G_{\mbox{\scriptsize pt-sm} } (s,t)$ denotes the standard
smeared-to-point propagator. In this form, the dependence on $\Delta_{T} $
resides purely in the (modified) propagators, and the derivative of the
correlator with respect to $\Delta_{T} $ can be assembled once one has
constructed the derivatives of the propagators. The derivative of a
modified point-to-point propagator $G$ is discussed in detail in
\cite{rome}; essentially, the derivative, evaluated at $\Delta_{T} =0$,
generates a vector current insertion into the propagator. Generalized
to smeared-to-point propagators, which contain an additional convolution
with a smearing kernel $K$, one obtains a further term from the derivative
of $K$ \cite{nhasan},
\begin{equation}
\left. \frac{\partial }{\partial q_j } G_{\mbox{\scriptsize pt-sm} } (s,t,q)
\right|_{q=0} = \sum_{x} G(s,x) \left[ -i \sum_{y} V_j G(x,y) K(y,t)
+ \left. \frac{\partial }{\partial q_j } e^{-iq\cdot (x-t)} K(x,t)
\right|_{q=0} \right]
\end{equation}
where $V_j $ is the conserved vector current insertion operator. The
computation of the derivative of the smearing kernel is discussed in
detail in \cite{nhasan}. By constructing and evaluating correlators
corresponding directly to derivatives of $C_3 $, cf.~(\ref{3ptcorr}),
in this fashion, any systematic bias in carrying out the derivative
with respect to $\Delta_{T} $ in (\ref{rdiscrete}) is avoided.

\begin{figure}[b]
\includegraphics[angle=-90,width=7.35cm]{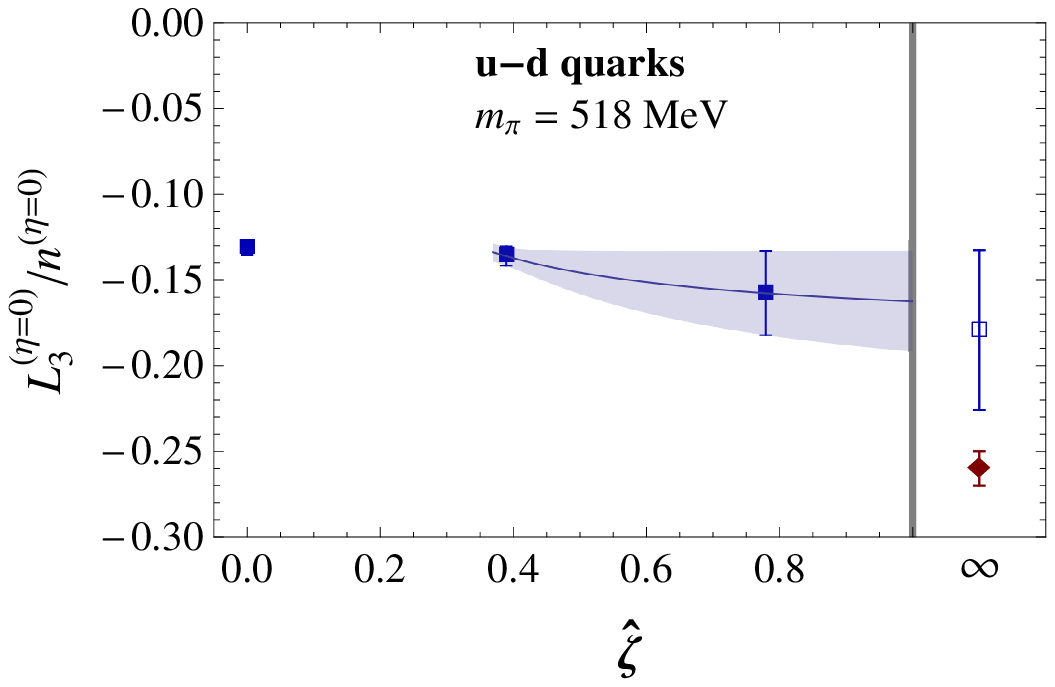} \\
\vspace{-4.83cm}

\hspace{8.2cm}
\includegraphics[width=6.85cm]{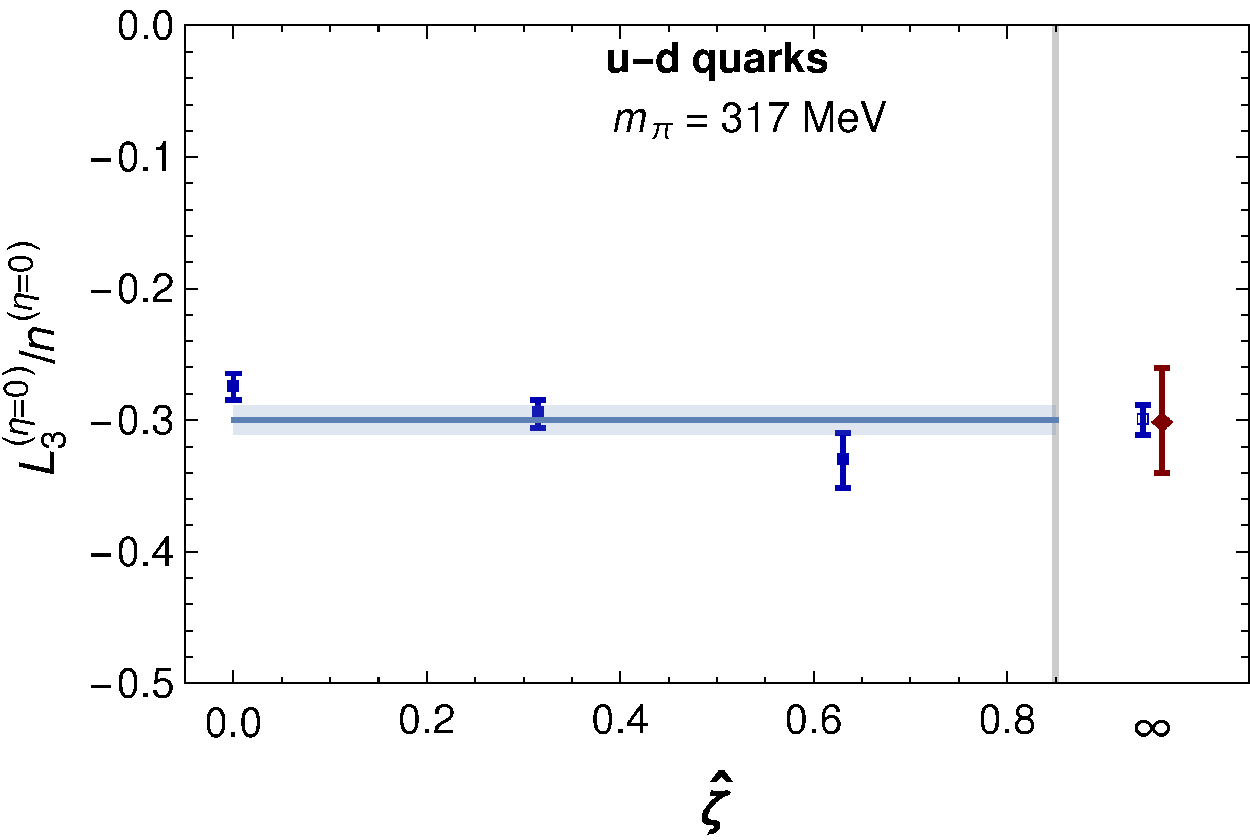}
\caption{Ji quark OAM obtained for different $\hat{\zeta } $, with
extrapolations to large $\hat{\zeta } $ (blue squares), compared to the
Ji sum rule result (red diamonds). The left panel is taken from
\cite{f14pap}, the right panel displays the results of the present work.}
\label{fig2}
\end{figure}

\section{Lattice calculation and results}
To perform a lattice calculation of the ratio (\ref{rdiscrete}), the
problem must be boosted into a Lorentz frame in which the TMD operator
entering $\Phi (z_T) $ exists at a single time. There is no obstacle to
this, given that the directions of $z$ and $v$ are both spacelike, cf.~the
discussion above preceding eq.~(\ref{ndenom}). In the frame
preferred for the lattice calculation, $v$ points in the longitudinal
3-direction, whereas $z_T $ is transverse, in the direction orthogonal
to the momentum transfer $\Delta_{T} $. In this frame, $\Phi (z_T) $ can
be evaluated using standard Lattice QCD methods. Numerical data for the
ratio (\ref{rdiscrete}) were obtained on a clover fermion ensemble
constituted of $32^3 \times 96$ lattices with spacing $a=0.114\, \mbox{fm} $
and pion mass $m_{\pi } =317\, \mbox{MeV} $.\hspace{0.09cm} The source-sink
separation employed was $10a = 1.14\, \mbox{fm} $.\hspace{0.09cm} The
longitudinal
\vspace{-0.4cm}

\begin{figure}[h]
\includegraphics[width=7.35cm]{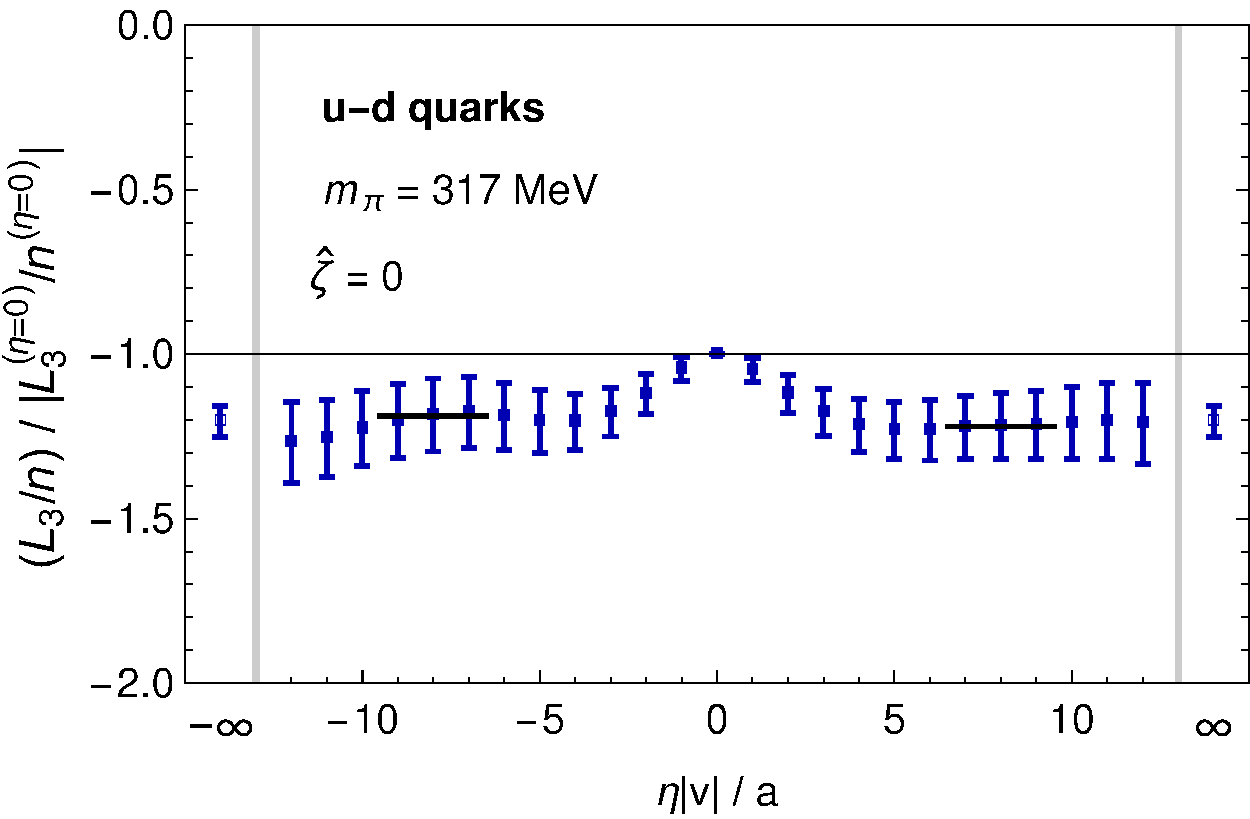} \\
\includegraphics[width=7.35cm]{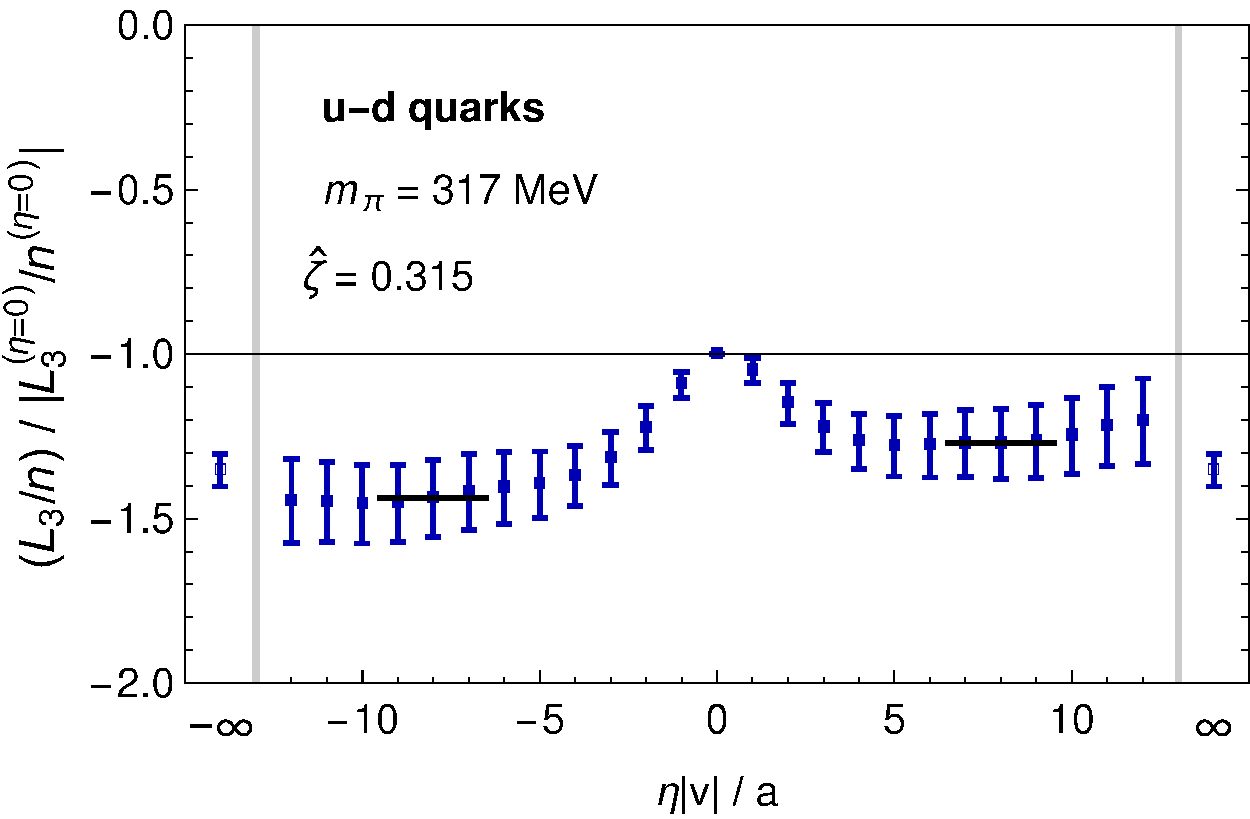} \\
\includegraphics[width=7.35cm]{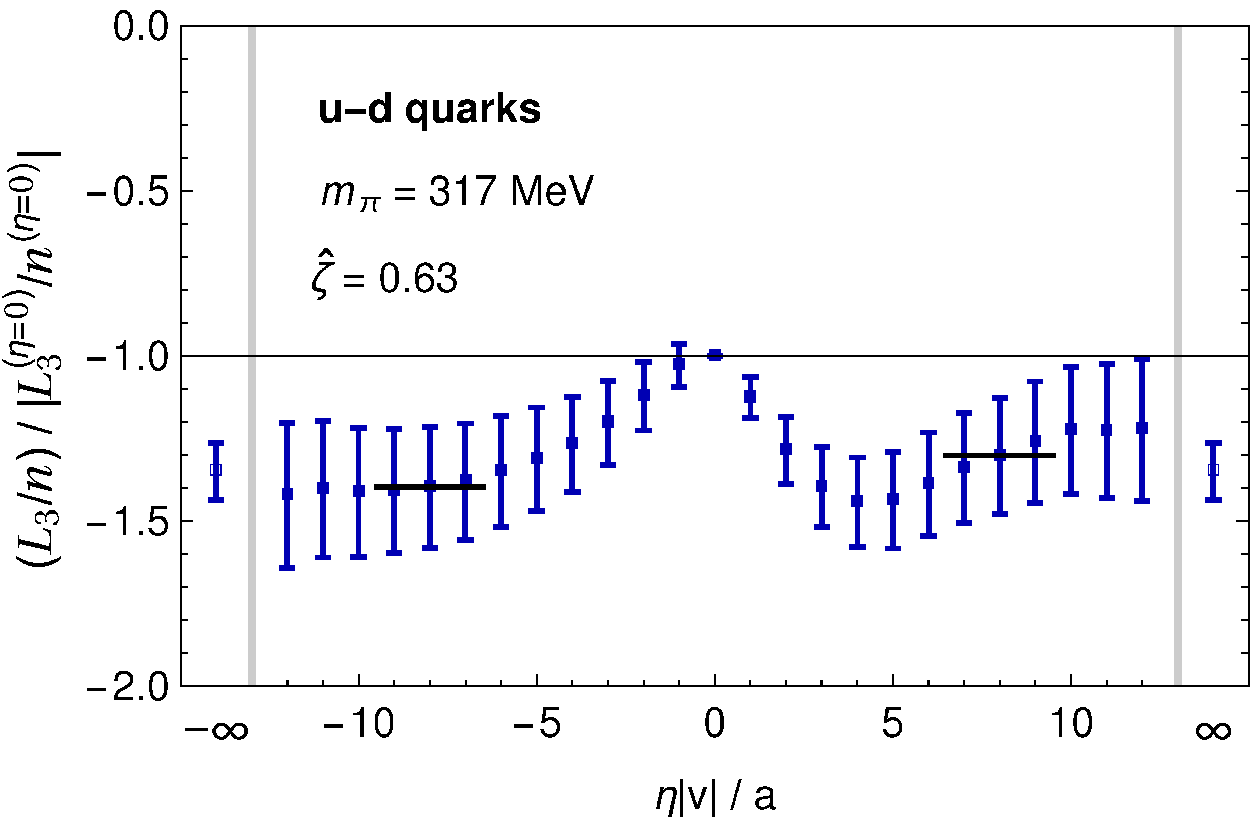}
\caption{Quark OAM as a function of staple length $\mbox{\hspace{7.45cm} } $
$\eta $, normalized to the modulus of the $\eta =0$ Ji OAM
$\mbox{\hspace{7.43cm} } $ value. Asymptotic values were
extracted by averaging $\mbox{\hspace{7.5cm} } $
over data at $\eta |v|/a=\pm 7$, $\pm 8$, $\pm 9$.
}
\label{fig3}
\vspace{-17.8cm}
\end{figure}

\hspace{7.5cm} \parbox{6.8cm}{
proton momentum
components $P_3 = 0$, $2\pi /(aL),\, 4\pi /(aL)$ were included in the
calculation, where $L=32$ denotes the spatial lattice extent. This
corresponds to Collins-Soper parameters $\hat{\zeta } = 0,\, 0.315,\, 0.63$.

\hspace{0.6cm}
Examining first the $\eta =0$ limit, corresponding to Ji OAM,
Fig.~\ref{fig2} compares the results obtained in the present
calculation (right panel) with results taken from the initial exploration
\cite{f14pap} (left panel). Data for three values of $\hat{\zeta } $ are
displayed, together with an extrapolation to large $\hat{\zeta } $,
compared to Ji OAM as obtained via Ji's sum rule at the same pion mass
\cite{LHPC}. It should be noted that, by maintaining the definition of
$v$ as pointing in the longitudinal 3-direction, $\hat{\zeta } $ can still
formally be defined in the $\eta =0$ case, and it characterizes the
momentum of the proton; however, Ji OAM ultimately cannot depend on
this parameter, since $v$ does not enter the definition of the straight
gauge link. I.e., Ji OAM is boost-invariant.

\hspace{0.6cm}
This constraint was not taken into account in the extrapolation of the
data in the left panel \cite{f14pap}, which themselves are compatible with
constant behavior. Instead, an ad hoc fit allowing for an approach to large
$\hat{\zeta } $ proportional to $1/\hat{\zeta } $ was performed. As a
result, the central value and uncertainty of the extrapolation extend
to larger magnitudes than they would by fitting a constant; with the latter
fit, the discrepancy with the Ji sum\hspace{0.105cm} rule\hspace{0.105cm}
value\hspace{0.105cm} would\hspace{0.105cm} be\hspace{0.105cm}
exhibited\hspace{0.105cm} more
}

\begin{figure}[t]
\includegraphics[width=7.35cm]{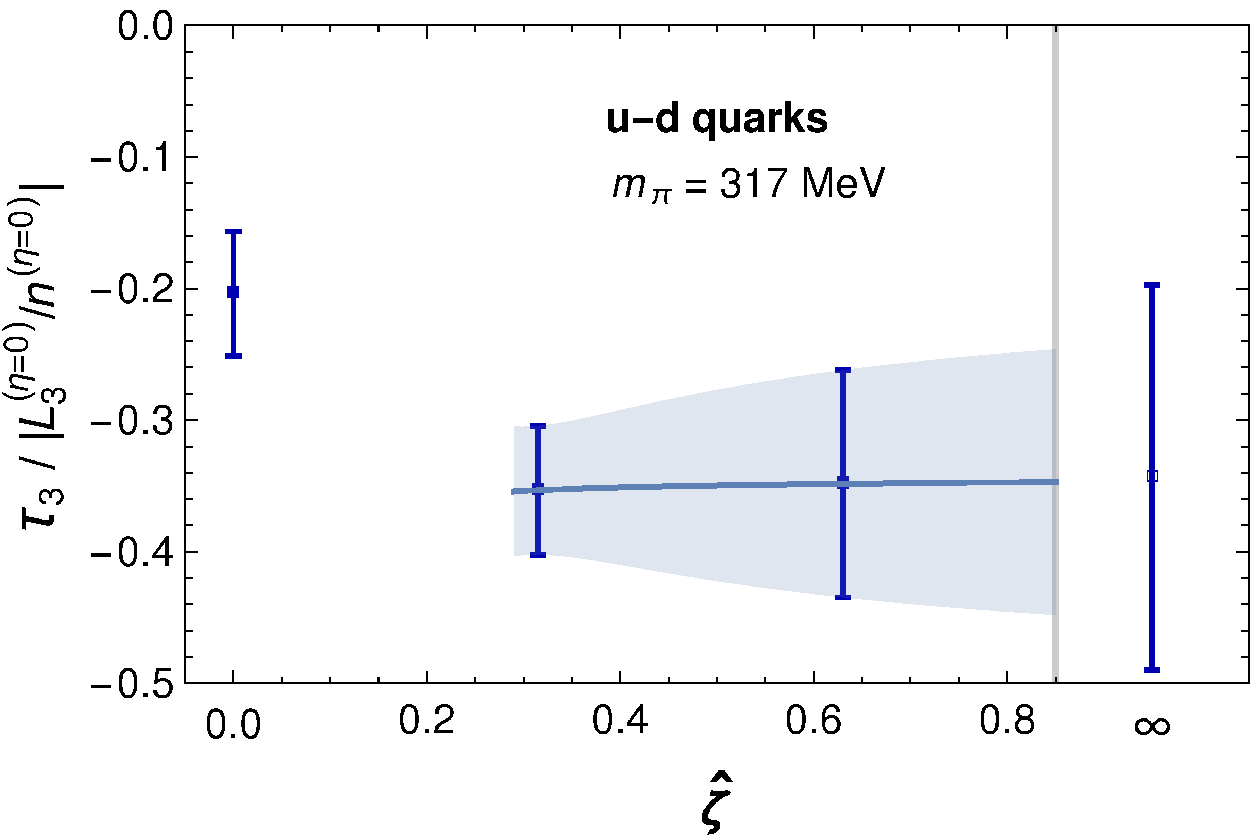}
\caption{Torque accumulated by the struck quark, cf.
$\mbox{\hspace{7.5cm} } $ main text, normalized to the modulus
of the $\eta =0$ Ji $\mbox{\hspace{7.43cm} } $ OAM value.}
\label{fig4}
\vspace{-7.4cm}
\end{figure}

\hspace{7.5cm} \parbox{6.8cm}{
starkly.
As already noted above, this discrepancy is owed to the biased estimate of
the $\Delta_{T} $-derivative in (\ref{rdiscrete}) via a finite difference
employed in \cite{f14pap}.

\hspace{0.6cm}
By contrast, the present calculation, cf.~the right panel in Fig.~\ref{fig2},
treats the $\Delta_{T} $-derivative in (\ref{rdiscrete}) in an unbiased
manner, and is seen to agree well with the Ji sum rule result for all
three proton momenta, as well as when extrapolated to large $\hat{\zeta } $
using a constant fit. This verifies that the discrepancy observed in the
initial study \cite{f14pap} indeed was due to the biased estimate of the
$\Delta_{T} $-derivative.\hspace{0.07cm} The\hspace{0.07cm}
unbiased\hspace{0.07cm} treatment\hspace{0.07cm} via\hspace{0.07cm} the
}
\vspace{-0.04cm}

\noindent
direct derivative method removes the discrepancy. The evaluation of Ji OAM
via the nonlocal GTMD matrix element (\ref{ldersingle}) coincides with the
evaluation via the local matrix elements encoding the GPD moments entering
Ji's sum rule, validating the GTMD method.

Turning to the transition from Ji OAM to Jaffe-Manohar OAM as a function
of the staple length $\eta $, Fig.~\ref{fig3} displays data at the three
different available $\hat{\zeta } $. The data are normalized to the
magnitude of the $\eta =0$ Ji value. Starting with Ji quark OAM at $\eta =0$,
the struck quark in a deep inelastic scattering process accumulates torque
as it is leaving the proton, to finally end up with Jaffe-Manohar OAM at
large $\eta $. The effect is substantial, can be clearly resolved in the
data, and is directed such as to enhance the magnitude of OAM compared to
the $\eta =0$ value. It increases as one departs from the $\hat{\zeta } =0$
limit towards finite proton momenta; no significant difference between
the results at the two nonvanishing $\hat{\zeta } $ is seen. The effect
is thus likely to survive the extrapolation to large $\hat{\zeta } $.
Fig.~\ref{fig4} displays such an extrapolation for the integrated torque
$\tau_{3} = L_{3}^{(\eta =\infty )} /n^{(\eta =\infty )} -
L_{3}^{(\eta =0)} /n^{(\eta =0)} $ alone, using the fit ansatz
$A+B/\hat{\zeta } $. The extrapolated integrated torque is roughly
one third of the originally present Ji quark OAM in the ensemble
used here.

% Figs.~\ref{fig2}-\ref{fig4} show results specifically for the isovector
% $u-d$ quark combination, in which disconnected contributions to
% $\Phi (z_T) $ exactly cancel. Fig.~\ref{fig5} exhibits flavor-separated
% data at $\hat{\zeta } =0.315$, along with the isoscalar $u+d$ total quark
% OAM, analogous to Fig.~\ref{fig3}. In these data, disconnected contributions
% are omitted; at the moderately high pion mass $m_{\pi } = 317\, \mbox{MeV} $
% considered in this work, they are expected to be suppressed compared to
% the connected contributions. The flavor-separated data show that the
% well-known cancellation between $u$- and $d$-quark OAM \cite{LHPC}
% persists away from the $\eta =0$ Ji limit.

% \begin{figure}[t]
% \includegraphics[width=7.35cm]{etaplot_upd_zeta315.ps}
% \caption{Caption ...
% }
% \label{fig5}
% \end{figure}

\section{Conclusion}
The main thrust of the work presented here was the further methodological
development of the GTMD approach to evaluating quark OAM in the proton
in Lattice QCD. Specifically, eq.~(\ref{rdiscrete}) calls for a
derivative with respect to momentum transfer $\Delta_{T} $. Employing
a direct derivative method to evaluate (\ref{rdiscrete}) free of
systematic bias, the result for Ji quark OAM was seen to agree with
the result obtained using the standard Ji sum rule method. This stands
in contrast to the initial exploration \cite{f14pap}, in which a biased
evaluation of the $\Delta_{T} $-derivative led to a significant
discrepancy in the Ji quark OAM results. The agreement achieved using
the improved methodology validates the GTMD approach. Furthermore, by
varying the gauge connection in the quark bilocal operator under
consideration, also Jaffe-Manohar OAM was extracted, and seen to be
enhanced significantly compared to Ji OAM, by about one third,
at the pion mass $m_{\pi } = 317\, \mbox{MeV} $ employed in this study.
Going forward, the exploration of quark OAM evolution through calculations
at varying lattice spacings is of interest, and investigations at lower
pion masses must be pursued.

\section*{Acknowledgments}
This work benefited from fruitful discussions with M. Burkardt,
S. Liuti and B. Musch. Computations were performed using resources
provided by the U.S.~DOE Office of Science through the National Energy
Research Scientific Computing Center (NERSC), a DOE Office of Science
User Facility, under Contract No.~DE-AC02-05CH11231, as well as through
facilities of the USQCD Collaboration at Fermilab, employing the Chroma
\cite{chroma} and Qlua software suites. R.~Edwards, B.~Jo\'{o}
and K. Orginos are acknowledged for providing the clover ensemble
analyzed in this work, which was generated using resources provided by
XSEDE (supported by National Science Foundation Grant No.~ACI-1053575).
S.M.~is supported by the U.S.~DOE, Office of Science, Office of High
Energy Physics under Award Number DE-SC0009913. S.S.~and S.M.~also
acknowledge support by the RHIC Physics Fellow Program of the RIKEN BNL
Research Center. M.E., J.N., and A.P.~are supported by the U.S.~DOE, Office
of Science, Office of Nuclear Physics through grants numbered
DE-FG02-96ER40965, DE-SC-0011090 and DE-FC02-06ER41444 respectively.
This work was furthermore supported by the U.S.~DOE through the TMD
Topical Collaboration.

\end{document}